\documentclass[prl,twocolumn,showpacs,amsmath,amssymb,citeautoscript]{revtex4}

 \def\be{\begin{eqnarray}}
\def\ee{\end{eqnarray}}

\usepackage{graphicx}

\begin{document}

\title{Anomalous superfluidity in $2+1$ dimensional two-color lattice QCD}
\author{Shailesh Chandrasekharan}
\affiliation{
Department of Physics, Box 90305, Duke University,
Durham, North Carolina 27708.}

\begin{abstract}
We study thermodynamics of strongly coupled lattice QCD with $two$ colors of staggered fermions in $(2+1)$ dimensions. The partition function of this model can be written elegantly as a statistical mechanics of dimers and baryonloops. The model is invariant under an $SO(3)\times U(1)$ symmetry. 
At low temperatures we find evidence for superfluidity in the $U(1)$ symmetry sector while the $SO(3)$ symmetry remains unbroken. The finite temperature phase transition appears to belong to the Kosterlitz-Thouless universality class, but the superfluid density jump $\rho_s(T_c)$ at the critical temperature $T_c$ is anomalously higher than the normal value of $2 T_c/\pi$. We show that by adding a small $SO(3)$ symmetry breaking term to the model, the superfluid density jump returns to its normal value implying that the extra symmetry causes anomalous superfluid behavior. Our results may be of interest to researchers studying superfluidity in spin-1 systems.
\end{abstract}

\maketitle

Two dimensional field theories are dominated by infrared fluctuations which lead to interesting non-perturbative phenomena. For example consider the $O(N)$ non-linear sigma model in two dimensions : When $N>2$ the theory is known to be asymptotically free and develops a non-perturbative mass gap, while the $N=2$ theory has both a critical (superfluid) phase and a massive (normal) phase, separated by the well known Kosterlitz-Thouless phase transition \cite{BKT2}. The transition is topological in nature and is driven due to the unbinding of topological defects (vortices). One of the striking theoretical predictions is that the superfluid density $\rho_s(T_c)$, at the critical temperature $T_c$, is known to jump from $2 T_c/\pi$ to zero \cite{Nelson} and has been observed in superfluid helium \cite{Bishop} and recently in spinless atomic BEC \cite{Hadzibabic}.

 Recently theories with an $SO(3)\times U(1)$ symmetry have become interesting due to the possibility of creating Bose condensates of atoms with non-zero spin, such as $^{23}Na$, $^{39}K$ and $^{87}Rb$, in quasi-two dimensional atomic traps \cite{Stock}. It is interesting to understand if these exotic spin systems can show new features. In a recent work it was argued that due to the presence of {\em half vortices}, superfluidity arising due to the condensation of spin-1 particles will be anomalous and in particular the superfluid density jump at $T_c$ will be $8 T_c/\pi$ instead of $2 T_c/\pi$ \cite{Muk06}. Some evidence for this conjecture was presented using Monte Carlo calculations on lattices as large as $40^2$. The Hamiltonian used to discuss the spin-1 spinor condensation was first introduced in \cite{Ho98}, where it was argued that the ground state can either be in a ``polar'' state or a ``ferromagnetic'' state. The fundamental difference between these two ground state manifolds leads to a different vortex structure such that superfluidity without spin order appears possible only in the polar phase \cite{Ho98,Zho01,Mak03,Muk06}. 

In this article we show that anomalous superfluidity also emerges naturally in a $2+1$ dimensional strongly coupled lattice gauge theory where fermions interact with an $SU(2)$ gauge field. As we will argue below, this model also has an $SO(3)\times U(1)$ symmetry and hence may be useful as an alternative model to study superfluidity in spin-1 systems. Fortunately, the partition function can be rewritten elegantly as a statistical mechanics of dimers and baryonloops making our model accessible to a variety of researchers. In this representation our model can be studied efficiently with ease on large lattices, thanks to recent developments in algorithms \cite{Adams:2003cc}. Here we report on results obtained from studies on lattices as large as $1024^2$. We find clear evidence for anomalous superfluidity: the jump in $\rho_s(T_c)$ is not close to its normal value of $2 T_c/\pi$. On the other hand, it is neither close to $8T_c/\pi$ as conjectured in \cite{Muk06} making our result even more interesting. Two-color lattice QCD (2CLQCD) has been studied in $3+1$ dimensions extensively as a toy model for the physics of strong interactions at finite temperatures and densities and continues to be interesting even today \cite{Hands:2006ve}. In the strong coupling limit it was studied recently using mean field theory \cite{Nishida:2003uj} and Monte Carlo methods \cite{Chandrasekharan:2006tz}. The theory has remained relatively unexplored in $(2+1)$ dimensions (however see \cite{Dunne:2003ji}).

The action of 2CLQCD that we study here is given by
\begin{equation}
S=\sum_{x\: {\rm even}}\frac{\eta_{\alpha,x}}{a_\alpha}
\Bigg[\overline{\Psi}_x U_{\alpha,x}\Psi_{x+\hat{\alpha}} - 
\overline{\Psi}_x U_{\alpha,x-\hat{\alpha}}^{\dagger}\Psi_{x-\hat{\alpha}}
\Bigg]
\label{act}
\end{equation}
where $x$ represents a site on a cubic lattice with coordinates $(x_t,x_1,x_2)$, $\overline{\Psi}_x=(\overline{\chi}(x),-\chi^{tr}(x)\tau_{2})$, and $\Psi^{tr}_x =(\chi^{tr}(x), \overline{\chi}(x)\tau_2)$ are four component Grassmann fields associated with even and odd lattice sites respectively. In our notation $\vec{\sigma}$ are Pauli matrices that mix $\chi$ and $\overline{\chi}^{tr}$ present in $\Psi$ and $\overline{\Psi}$ while $\vec{\tau}$ are Pauli matrices that act on the color space. The Grassmann valued quark fields $\overline{\chi}(x)$ and $\chi(x)$ represent row and column vectors with $2$ color components. The gauge fields $U_{\alpha,x}$ are elements of $SU(2)$ group and live on the links between $x$ and $x+\hat{\alpha}$ where $\alpha=t,1,2$. The factors $\eta_{t,x}=1$, $\eta_{1,x}=(-1)^{x_t}$ and $\eta_{2,x}=(-1)^{x_t+x_1}$ are the staggered fermion phase factors. We fix $a_1=a_2=1$ but vary $a_t$. By choosing a slab geometry of the type $L_t\times L^2$ (periodic in all directions) we can study thermodynamics in the $L \rightarrow \infty$ limit at a fixed $L_t$. Our results were obtained with $L_t=4$. The parameter $\delta=1/(a_t)^2$ then controls the temperature. 

The action (\ref{act}) is invariant under $\Psi \rightarrow V \Psi $, $\overline{\Psi} \rightarrow\overline{\Psi} V^{\dagger}$ where $V = \exp(i\vec{\phi}\cdot\vec{\sigma}+i\theta) \in U(2)$. Since $V=1$ when $\vec{\phi} = (0,0,\pi)$ and $\theta =\pi$ the real symmetry of the model is $SO(3)\times U(1)$, similar to the symmetry of the model studied in \cite{Muk06}. In two spatial dimensions, at zero temperatures, the above model is expected to develop a ``chiral condensate'' which means $\langle \Psi^{tr} \sigma_1 \Psi \rangle = \langle \overline{\Psi} \sigma_1 \overline{\Psi}^{tr} \rangle \equiv \langle \bar\chi\chi\rangle \neq 0$. Clearly this breaks the $SO(3)\times U(1)$ symmetry to $O(2)$. Note that the order parameter is invariant under $V = \exp(i\phi_3\sigma_3) (1,\sigma_1)$ which is simply isomorphic to $O(2)$. Thus, the Goldstone boson manifold is the same as that of the polar phase discussed in \cite{Mak03,Muk06}. This connection should make our work interesting to researchers studying spin-1 spinor condensates.

\begin{figure}
\begin{center}
\includegraphics[width=0.4\textwidth]{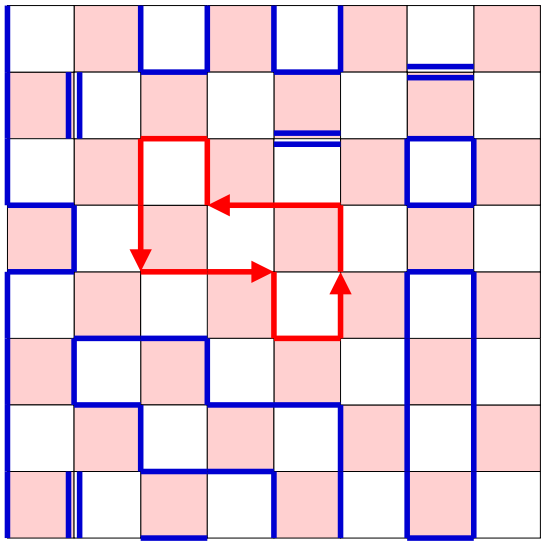}
\end{center}
\caption{\label{fig1} An illustration of a dimer-baryonloop configuration in $(1+1)$ dimensions.}
\end{figure}

It is possible to integrate out the gauge fields exactly and write the partition function as a statistical mechanics of dimers with constraints \cite{Rossi:1984cv,Chandrasekharan:2006tz}. A dimer is a bond variable connecting two neighboring sites $x$ and $x+\hat{\alpha}$. There are two types of dimers possible in our model: an oriented dimer (with an arrow) and a dimer without orientation. Oriented dimers will be represented by $b_\alpha(x)=1,-1$, while non-oriented dimers will be represented by $d_\alpha(x)=1,2$. If a particular type of dimer is absent then the corresponding $b_\alpha(x)$ or $d_\alpha(x)$ is set to zero. Clearly $d_\alpha(x) = d_{-\alpha}(x+\hat{\alpha})$ and $b_\alpha(x) = -b_{-\alpha}(x+\hat{\alpha})$. The microscopic model also introduces two constraints: (1) the sum of $d_\alpha(x) + |b_\alpha(x)|$ over all the bonds $\alpha$ connected to a site $x$ must be $2$, (2) every site must have one incoming and one outgoing oriented dimer or no oriented dimers.   Thus, each configuration consists of three types of connected objects: (a) self avoiding oriented loops made with oriented dimers (called baryonloops), (b) self avoiding non-oriented loops made with non-oriented dimers, and (c) double dimer bonds. For illustration we show an example of a $(1+1)$ dimensional dimer-baryonloop configuration in figure \ref{fig1}. However, note that our work is done in $(2+1)$ dimensions.

The Boltzmann weight of each configuration is given by $\delta^n$ where $n$ is the total number of temporal dimers (oriented or otherwise). We focus on two types of winding number susceptibilities:
\begin{subequations}
\begin{eqnarray}
\chi_w^{u(1)} &=& \frac{1}{2 L^2}\Bigg\langle 
\Bigg(\sum_{x,\alpha=1,2} J_\alpha^{u(1)}(x)\Bigg)^2 \Bigg\rangle,
\\
\chi_w^{so(3)} &=& \frac{1}{2 L^2} \Bigg\langle 
\Bigg(\sum_{x,\alpha=1,2} J_\alpha^{so(3)}(x) \Bigg)^2 \Bigg\rangle,
\end{eqnarray}
\end{subequations}
constructed from the conserved currents $J_\alpha^{u(1)}(x) = \sigma_x [|b_\alpha(x)|+d_\alpha(x) - 2]$, and $J_\alpha^{so(3)} = b_\alpha(x)$ which arise due to the $U(1)$ and $SO(3)$ symmetries discussed above. Here, $\sigma_x=1$ on even sites and $-1$ on odd sites.
 
\begin{figure}
\begin{center}
\includegraphics[width=0.45\textwidth]{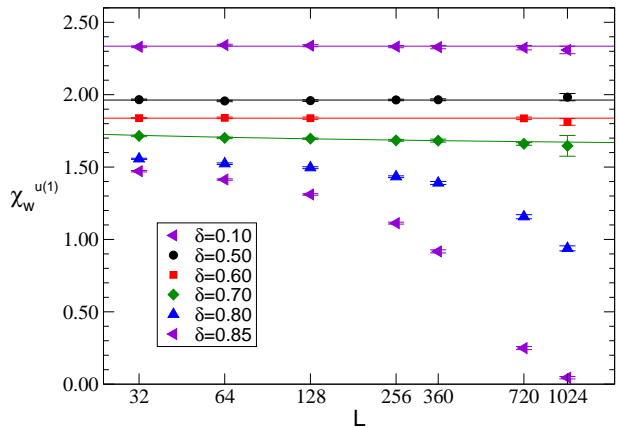}
\end{center}
\caption{\label{fig2} Plot of $\chi_w^{u(1)}$ versus $L$ for different values of $\delta$. At $\delta=0.1,0.5$ and $0.6$ we find $\chi_w^{u(1)} = 2.335(3), 1.963(3)$ and $1.838(3)$ fits the data well (solid lines). At $\delta=0.7$ eq.(\ref{ktwind}) with $A=1.59(1)$ and $B=2.7(7)$ fits better.(dashed line).}
\end{figure}

In the superfluid phase we expect $\chi_w^{u(1)} \neq 0$ for large $L$. In figure \ref{fig2} we plot $\chi_w^{u(1)}$ as a function of $L$ for different values of $\delta$. For $\delta \leq 0.6$, $\chi_w^{u(1)}$ is independent of $L$ in the range $32 \leq L \leq 1024$, suggesting that we are in the superfluid phase. When $\delta \geq 0.8$ we see that $\chi_w^{u(1)}$ begins to decrease for large $L$ suggesting that superfluidity is lost. For $\delta = 0.7$ our data fits well to the form
\begin{equation}
\chi_w^{u(1)} = A \Bigg[1 + \frac{0.5}{\log(L)+B}\Bigg]
\label{ktwind}
\end{equation}
expected to hold at the KT transition. We find $A=1.59(1)$ and $B=2.7(7)$, with a $\chi^2/DOF = 0.5$. Since $\chi_w^{u(1)} = \rho_s(T)/T$ \cite{mismatch}, one would have expected $A$ to be close to either $2/\pi$ for normal superfluidity or $8/\pi$ for superfluidity due to half vortices \cite{Muk06}. Our data is inconsistent with both these expectations.

\begin{figure}
\vskip0.2in
\begin{center}
\includegraphics[width=0.45\textwidth]{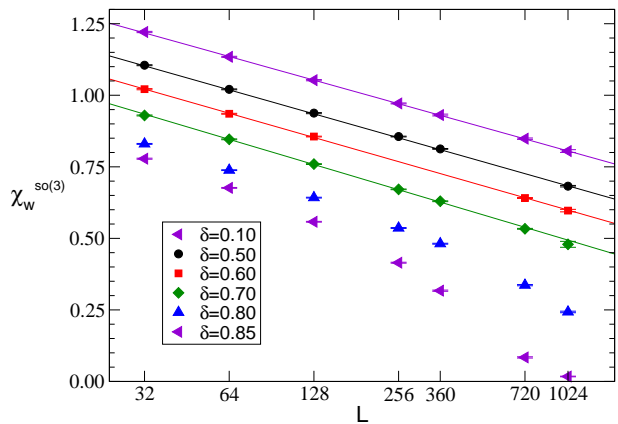}
\end{center}
\caption{\label{fig3} Plot of $\chi_w^{so(3)}$ versus $L$ for different values of $\delta$. The solid lines are fits to eq.(\ref{so3fit}) discussed in the text.}
\end{figure}

In two dimensions there is no spontaneous symmetry breaking which means we should expect $\chi_w$ to go to zero for large $L$. Of course $U(1)$ symmetry is special which allows for a non-zero $\chi_w^{u(1)}$, but $\chi_w^{so(3)}$ must vanish for large $L$. If $\xi$ is the correlation length in the $SO(3)$ channel, we expect $\chi_w^{so(3)}$ is exponentially small for $L \gg \xi$. However, for small values of $\delta$ our results are obtained in the region where $L\ll \xi$ since $\xi$ is expected to be asymptotically large. Here the $L$ dependence of $\chi_w^{so(3)}$ can be found by solving a one-loop renormalization group equation \cite{Azaria} which gives
\begin{equation}
\chi_w^{so(3)}= C - D\log(L).
\label{so3fit}
\end{equation}
Our results for $\delta < 0.7$ are consistent with this behavior while at $\delta=0.7$ it just begins to break down suggesting that $\xi \sim 1024$ there. The fits are as follows:
\begin{center}
\begin{tabular}{|c|c|c|c|}
\hline
$\delta$ & $C$ & $D$ & $\chi^2/DOF$ \\
\hline
0.1 & 1.632(4) & 0.119(1) & 0.47 \\
0.5 & 1.526(3) & 0.120(1) & 0.51 \\
0.6 & 1.443(4) & 0.122(1) & 0.18 \\
0.7 & 1.373(3) & 0.127(1) & 5.6 \\
\hline
\end{tabular}
\end{center}
For larger values of $\delta$, $\chi_w^{so(3)}$ begins to decrease faster.
These results are plotted in figure \ref{fig3}.

Based on the above analysis we conclude that in our model the $U(1)$ symmetry leads to anomalous superfluidity at low values of $\delta$ while the $SO(3)$ symmetry remains unbroken. We estimate that $\delta_c$ is roughly between $0.6$ and $0.8$. An obvious criticism to our conclusions is the following : Perhaps we have not reached the thermodynamic limit for $\delta \sim 0.6$ since in this region $L \ll \xi$. Is it not possible that $\chi_w^{u(1)}$ remains almost a constant until $L \sim \xi$ and then begins to drop to a different value \cite{windcomm}? If true, this would imply that $\delta_c$ could be much smaller than our estimate so that the jump in $\chi_w^{u(1)}$ at $\delta_c$ could indeed be $8/\pi$.

\begin{figure}
\vskip0.2in
\begin{center}
\includegraphics[width=0.45\textwidth]{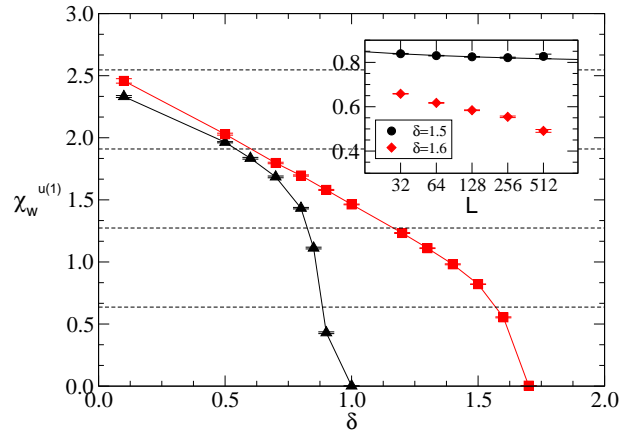}
\end{center}
\caption{\label{fig4} Plot of $\chi_w^{u(1)}$ versus $\delta$ at $L=256$ for $W=1$ (triangles) and $W=0.9$ (squares). The solid lines just connect the data points to guide the eye. The inset shows the finite size scaling plot similar to figure \ref{fig2} at $W=0.9$. The values $2\nu/\pi,\nu =1,2,3,4$ are shown as dashed lines for reference.}
\end{figure}

If the above criticism is true, one would expect that $\chi_w^{u(1)}$ is very sensitive to $\xi$ in the region $\delta \sim 0.6$. We should be able to study this sensitivity by changing $\xi$. We can do this by changing the Boltzmann weight of each dimer-baryonloop configuration to $\delta^n W^m$ where $m$ is the number of oriented dimers in the configuration. When $W=1$ we obtain the original model, but when $W < 1$, $\xi$ is considerably shortened since the $SO(3)$ symmetry of the original model is explicitly broken to a $U(1)$ subgroup \cite{wcomment}. Changing $W$ does not affect the superfluid $U(1)$ symmetry. The two currents $J_\alpha^{so(3)}$ and $J_\alpha^{u(1)}$ continue to be conserved due to the remnant $U(1)\times U(1)$ symmetry of the modified model.

The $W=0$ model was studied in \cite{Chandrasekharan:2003qv} where it was shown that $\chi_w^{u(1)}$ shows the normal $2/\pi$ jump at the KT transition. When $W=0$ there are no oriented dimers which means that $\chi_w^{so(3)}=0$ exactly. In this work we have studied the modified model for $W=0.9$ and we again find that $\chi_w^{so(3)}=0$ for the lattice sizes we have studied showing again that $\xi$ has shortened considerably. In figure \ref{fig4} we plot $\chi_w^{u(1)}$ as a function of $\delta$ for $L=256$. For comparison we plot the values for both $W=0.9$ and $1$. The inset of figure \ref{fig4} shows the dependence of $\chi_w^{u(1)}$ on $L$ at $\delta=1.5$ and $1.6$ for $W=0.9$. The data at $1.5$ fits to the form given in eq.(\ref{ktwind}), with $A=0.771(4)$, $B=2.2(4)$ with $\chi^2/DOF=0.7$ while at $1.6$ the fit is poor. This suggests $1.5 < \delta_c < 1.6$ and the jump in $\chi_w^{u(1)}$ at $\delta_c$ is now very close to the normal value of $2/\pi$.

Clearly, the breaking of the $SO(3)$ symmetry has a significant effect on the critical point and the jump in the superfluid density at the critical point returns to its normal value. However, changing $W$ from $1$ to $0.9$ has negligible effect on $\chi_w^{u(1)}$ in the region $\delta \sim 0.6$. This weakens significantly the criticism raised above that our lattices are far from the thermodynamic limit for $\delta \sim 0.6$. 


\begin{figure}
\vskip0.2in
\begin{center}
\includegraphics[width=0.45\textwidth]{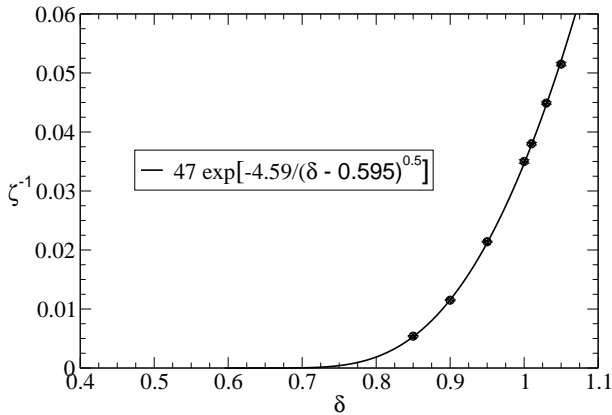}
\end{center}
\caption{\label{fig5} Plot of $\zeta^{-1}$ (extrapolated to the infinite volume) versus $\delta$ for $W=1$.}
\end{figure}


We have also measured the correlation length $\zeta$ using the correlator $\langle \overline{\chi}_x \chi_x \quad \overline{\chi}_y \chi_y \rangle$ where $x$ and $y$ are two lattice sites. We look for an exponential decay along a spatial direction while projecting to zero momentum (or frequency) in the other directions. Figure \ref{fig5} shows the behavior of $\zeta^{-1}$ extracted from our data after removing all finite size effects. We see that $\zeta$ appears to be diverging rapidly. However, note that the correlator we measure couples to fluctuations which have both $SO(3)$ as well as $U(1)$ quantum numbers. Hence it is possible that $\zeta$ grows asymptotically large as $\delta$ is lowered but does not diverge at the critical point. Clearly our data cannot be used to confirm this. However, in order learn something, here we fit the data to the form $\zeta^{-1} = \alpha \exp(-\beta/\sqrt{\delta-\delta_c})$, expected at a $KT$ transition. We find $\alpha = 47(15)$, $\beta=4.59(3)$ and $\delta_c \sim 0.595(20)$ with a $\chi^2/DOF=0.85$. Interestingly this estimate of $\delta_c$ is not very different from our previous estimates.

Before we end we would like to note that it is possible to modify our model so that the jump at $T_c$ is $8 T_c/\pi$ as predicted in \cite{Muk06}. For example if the Boltzmann weight of each dimer-baryonloop configuration is chosen to be $\delta^n (W_D)^m$, where $m$ is the number of $d_\alpha(x)=2$ bonds in the configuration, then it is easy to argue that the $SO(3)\times U(1)$ symmetry remains unaffected. For large $W_D$ all loops will be suppressed and the model should behave like the close-packed dimer model studied in \cite{Chandrasekharan:2003qv}. The only difference is that the current $J_\alpha^{u(1)}$ will now be twice the value in the earlier study. Since the earlier study showed a normal jump in the superfluid density, our model with large $W_D$ should show four times the normal jump. We have confirmed this for $W_D=3.5$. This is not surprising since increasing $W_D$ makes the $SO(3)$ symmetry more disordered and thus makes half-vortex excitations easy. It would be interesting to study how the superfluid density jump evolves with $W_D$ and understand if an interplay between half-vortices and normal vortices can lead to the anomalous superfluid behavior observed here.


\section*{Acknowledgments}

We thank H.~Baranger, F.-J.Jiang, A.~Vishwanath, S.~Sachdev and E.~Vicari for helpful comments. This work was supported in part by the DOE grant DE-FG02-05ER41368 and NSF grant DMR-0506953. Parts of this work first appeared in \cite{Chandrasekharan:2005dn}.

\bibliography{qcd,footnotes}

\begin{thebibliography}{21}
\bibitem[{mis()}]{mismatch}
\bibinfo{note}{Here we ignore the small renormalization effects that occur due
  to the choice of $L_1=L_2$.}
\bibitem[{win()}]{windcomm}
\bibinfo{note}{Physically it seems unlikely that $\chi_w^{u(1)}$ can be larger
  for $L\gg\xi$ compared to $L\ll\xi$.}
\bibitem[{wco()}]{wcomment}
\bibinfo{note}{Without breaking the SO(3) symmetry we find it difficult to
  reduce $\xi$.}
\end{thebibliography}

\end{document}